\documentclass[prl,aps,twocolumn,groupedaddress,showpacs,floatfix]{revtex4}

\usepackage{amsmath,amssymb,multirow,epsfig,bm}

\newcommand{\beq}{\begin{equation}}
\newcommand{\eeq}{\end{equation}}
\newcommand{\bea}{\begin{eqnarray}}
\newcommand{\eea}{\end{eqnarray}}

\newcommand{\nn}{\nonumber}
\newcommand{\benn}{\begin{displaymath}}
\newcommand{\eenn}{\end{displaymath}}
\begin{document}

\title{\bf Neutron localization induced by the pairing field in an inhomogeneous neutron matter}

\author{Piotr Magierski}

\affiliation{Department of Physics, University of
Washington, Seattle, WA 98195--1560, USA}

\affiliation{Faculty of Physics, Warsaw University of Technology,
ul. Koszykowa 75, 00-662 Warsaw, POLAND }

\date{\today}

\begin{abstract}
It is shown that in an inhomogeneous neutron matter the pairing field 
bounds neutrons around the Fermi level leading to the formation of localized 
Andreev states. In the case of the inner crust of neutron stars
the localization length has been determined as a function of the nuclear density. 
\end{abstract}


\maketitle

The inhomogeneous nuclear matter forms the inner crust of neutron stars
at densities ranging from $10^{-4}$ fm$^{-3}$ to $10^{-1}$ fm$^{-3}$.
The nuclear matter is thus diluted and the nucleons are
strongly paired through the attraction in the s-wave channel.
Since the proton fraction does not exceed a few percent  
in this density region the main contribution to the pairing energy 
comes from neutrons. The dependence of the pairing field on the neutron density
has been a subject of an intensive theoretical studies 
(see \cite{dh} and references therein). 
Approaches based on the BCS theory predict the maximum of the s-wave pairing gap $\Delta$ 
to be around $3 MeV$\cite{kkc}, whereas the inclusion of polarization corrections decreases
this value to about $1 MeV$\cite{schulze}. Recently, the non-perturbative calculations
for fermions in the unitary limit predict the pairing gap at $T=0$
of about $0.6\epsilon_{F}$, where $\epsilon_{F}$ is the Fermi energy
\cite{carlson}. Since the neutron 
matter at the densities between $10^{-3}$ fm$^{-3}$ and $10^{-2}$ fm$^{-3}$
is close to the unitary limit, it suggests that the pairing gap is 
a considerable fraction of the Fermi energy.

The proton admixture drives the nuclear matter towards an inhomogeneous state
and results in the appearance of nuclear impurities immersed
in the superfluid neutron liquid. Inside impurities the density reaches
the nuclear saturation density.
The lattice of nuclei is stabilized by the Coulomb interaction and
becomes denser as the average density increases, and eventually
melts due to the interplay between Coulomb, surface and shell energies 
\cite{pra,bma,mph}.
Consequently, the inhomogeneous neutron density distribution induces the spatial
variations of the pairing potential\cite{brog,sand}. 
Because the structure of the major part of the crust is predicted
to be periodic, the pairing field is periodic as well.
Moreover pairing at nuclear saturation density is usually weaker
than in the dilute neutron matter 
(except for a very dilute system with a density smaller than $10^{-3}$fm$^{-3}$). Hence the 
local minima of the pairing field coincide with the position of nuclear impurities.
The scale of spatial variations of the pairing field is governed by the coherence length 
$\xi=\displaystyle{\frac{1}{k_{F}}\frac{\epsilon_{F}}{\Delta}}$, where
$k_{F}$ stands for Fermi momentum. Although the pairing gap is large, it is still
smaller than $\epsilon_{F}$ and thus the variations of the pairing field
are smoother than that of the nuclear mean-field potential.

The inhomogeneity of the pairing field induces an additional scattering 
of quasiparticles. Because the pairing
field is periodic, an additional Bragg scattering 
gives rise to corrections to the neutron band structure.
Therefore the natural question appears to be whether this effect is large, or can
be neglected. This article provides an answer to this question and discuss the impact 
of the inhomogeneity of the pairing field on the quasiparticle states.
It is organized as follows: first, I describe the mechanism of bound state formation
in the schematic model.
Second, it will be shown, that contrary to the normal three-dimensional potential,
in the case of a pairing well, there is always a bound state, no matter
how shallow the well is.
Finally, I will present the variational calculations for realistic pairing
potentials through the inner crust and estimate the density region where 
the localized states appear. 

In order to investigate the problem one must realize 
that the scattering of quasiparticles on the pairing field
is qualitatively different than that on the normal potential. Namely, the
scattering matrix contains in this case the normal part which comprises
the scattering amplitude of hole into hole and particle into particle, but
in addition it contains the amplitude of scattering of particle into hole and vice 
versa. This, so called Andreev reflection has an unusual property that
the particle/hole is reflected almost exactly in the same direction as
the incoming hole/particle \cite{ag}.
In fact, it can be shown that the Andreev reflection dominates around the Fermi level
and gives rise to the formation of the so-called Andreev states\cite{ag, bpw}. 

The spatial variations of the pairing field may be simulated
in the first approximation by
the simple model consisting of the spherical pairing well:
\beq
\Delta(r) = \Bigg{\{}
\begin{array}{cc}
\Delta_{in}  & \mbox{\hspace{0.5cm}for\hspace{0.5cm}} r < R  \\
\Delta_{out} & \mbox{\hspace{0.5cm}otherwise},
\end{array}
\eeq
and $|\Delta_{out}|>|\Delta_{in}|$. In the absence of superflow the
phase of the pairing field is constant and can be chosen to be zero.
The BdG equation takes the form:
\beq \label{bdg}
\left ( \begin{array}{cc} T -\mu & \Delta(r) \\
                          \Delta (r)& -(T-\mu)
\end{array} \right )
\left ( \begin{array}{l} u({\bf r}) \\   v({\bf r})
\end{array} \right )
= E \left ( \begin{array}{l} u({\bf r}) \\   v({\bf r})
\end{array} \right ),
\eeq
where $T$ is the kinetic energy operator (for simplicity
we disregard the normal potential), $E$ is the quasiparticle
energy, $u({\bf r})$ and $v({\bf r})$ are the particle and hole
amplitudes of the quasiparticle wave function.
In the simplest Andreev approximation, which will admit
the analytic solution, the wave function is decomposed
into the rapidly oscillating part, and the smooth part, related to
the changes of the pairing field \cite{andr}:
\beq \label{decomp}
\left ( \begin{array}{l} u({\bf r}) \\   v({\bf r})
\end{array} \right )
= \exp (i {\bf k_F}\cdot{\bf r} )
\left ( \begin{array}{l} \bar{u}({\bf r}) \\   \bar{v}({\bf r})
\end{array} \right ),
\eeq
where $\bar{u}({\bf r})$ and $\bar{v}({\bf r})$ changes slowly
within the scale determined by the coherence length.
Neglecting terms of the order of $1/(k_{F}\xi)$ the original equations 
transform into:
\beq \label{andreev}
\left ( \begin{array}{cc} -2 i {\bf k_{F}}\cdot\nabla -\mu & \Delta(r) \\
                          \Delta (r)& 2 i {\bf k_{F}}\cdot\nabla + \mu
\end{array} \right )
\left ( \begin{array}{l} \bar{u}({\bf r}) \\   \bar{v}({\bf r})
\end{array} \right )
= E \left ( \begin{array}{l} \bar{u}({\bf r}) \\   \bar{v}({\bf r})
\end{array} \right ).
\eeq
We have chosen units: $\frac{\hbar^{2}}{2m}=1$, i.e.
$T=-\nabla^{2}$ and $k_{F}^2 = \mu$.
Although the Andreev approximation must be used with care
for a very dilute neutron matter, it is instructive to study this limit
(i.e. $k_{F}\xi\rightarrow \infty$, $\Delta /\mu \rightarrow 0$),
as the consequences of this oversimplified approach remain valid for a more realistic cases.

The astonishing implication of eq. (\ref{andreev}) is that the three-dimensional
problem is transformed into a one dimensional, and
for each path specified by the vector $\bf{k_{F}}$
exists a solution. This result is a direct
consequence of the focusing property of the Andreev reflection\cite{ag,bpw}.
In order to investigate the $L=0$ state, one must
consider the quantization condition along the diameter:
\beq \label{quant}
A(\phi_{+},\phi_{-})\exp (2 i q R) = 1 ,
\eeq
where
\bea
A(\phi_{+},\phi_{-}) &=& \frac{(e^{-\phi_{-}} - e^{-i\phi_{+}} )( e^{\phi_{-}} - e^{i\phi_{+}}  )}
                            {( e^{-\phi_{-}} - e^{i\phi_{+}}  )( e^{\phi_{-}} - e^{-i\phi_{+}} )} , \nn \\
2 q R &=&  k_{F} R\frac{\Delta_{out}}{\mu} 
\sqrt{\left (\frac{E}{\Delta_{out}}\right )^2-\left (\frac{\Delta_{in}}{\Delta_{out}}\right )^2}
\eea
and
\bea
\cos(\phi_{+}) &=&\frac{E}{\Delta_{out}} ,\nn \\
\cosh(\phi_{-})&=&\frac{E}{\Delta_{in}}. \nn
\eea
The closer inspection of the above result reveals that the equation
(\ref{quant}) always admits a solution for $E$ in the interval
$(\Delta_{in},\Delta_{out})$, no matter how small
is the difference $|\Delta_{in}|-|\Delta_{out}|$. Namely, in the limit 
$|\Delta_{in}| < |\Delta_{out}| \rightarrow 0$ the quantization condition becomes:
\beq
\exp (2 i (q R + \Phi) ) = 1,
\eeq
where
\bea
\cos \Phi & = &  \frac{2y - x(1+y)}{x(1-y)} ,\nn \\
x  & =  & \frac{E}{\Delta_{out}} , \\
y  & =  & \frac{\Delta_{in}}{\Delta_{out}} \nn .
\eea
Since $qR\rightarrow 0$ and $y\le x\le 1$ it is easy to find that the solution reads:
\beq \label{limit1}
x \approx \frac{1}{1 + (qR)^2\frac{1-y}{y}} \le 1.
\eeq
Another limit relevant for the neutron star crust is when $\Delta_{in}\rightarrow 0$.
In that case one gets:
\beq \label{limit2}
\frac{1}{2}k_{F} R \frac{E}{\mu} - \arccos\frac{E}{\Delta_{out}} = \pi m, 
\mbox{\hspace{0.5cm}} m=0,\pm 1, \pm 2,...
\eeq
The examination of both limits reveals that once $\displaystyle{ k_{F} R \frac{\Delta_{out}}{\mu}}$ and
$\displaystyle{\frac{\Delta_{in}}{\Delta_{out}}}$ are fixed,
then $\displaystyle{\frac{E}{\Delta_{out}}}$
is constant as well. This property is worth mentioning in the context of the neutron
star crust, because there $R \propto \xi$, as the size of nuclear impurities is small
with respect to the coherence length. In this case clearly $\displaystyle{ k_{F} R \frac{\Delta_{out}}{\mu}}$
is approximately constant and thus $\displaystyle{\frac{E}{\Delta_{out}}}$ stays fixed as well,
through the whole inner crust. Consequently, one may expect that the size of the tail of the 
wave function bound by the pairing field, $\eta$, will be proportional to $\xi$ and $\eta/\xi$ will 
be approximately constant through the crust. In the following we will show that this is indeed
the case.

Let us turn now to the three-dimensional description, in order to study
whether the above results survive in a more realistic case.
In the variational formulation of the problem (\ref{bdg}), one must consider the 
expectation value of the square of the energy:
\beq \label{variational}
\langle E^2\rangle = \int d^{3} r \chi^{+}({\bf r})\hat{\Omega}^{+}\hat{\Omega} \chi({\bf r}),
\eeq
where
\beq
\hat{\Omega}=\left ( \begin{array}{cc} T -\mu & \Delta({\bf r}) ,\\
                          \Delta ({\bf r})& -(T-\mu),
\end{array} \right )
\eeq
and $T=-\nabla^{2}$ and $k_{F}^2 = \mu$.
We use the ansatz so that the wave function $\chi({\bf r})$ can be decomposed 
similarly to (\ref{decomp}):
\beq
\chi({\bf r})=\phi_{k_{F}}({\bf r})\psi_{k_{\eta}}({\bf r}),
\eeq
where $\psi_{k_{\eta}}({\bf r})$ is a spinor, 
and the following conditions hold:
\bea
-\nabla^{2}\phi_{k_{F}}({\bf r}) & = &  k_{F}^2\phi_{k_{F}}({\bf r}) , \nn \\
-\nabla^{2}\psi_{k_{\eta}}({\bf r})  & \propto  &  k_{\eta}^{2} \psi_{k_{\eta}}({\bf r}), \\
\frac{k_{\eta}}{k_{F}} & \ll  & 1 . \nn
\eea
Because the problem is time-reversal invariant then $\chi({\bf r})$ can be chosen
to be real, and consequently the term $\langle [\nabla^2, \Delta(r) ]\rangle =0$.
Hence (\ref{variational}) can be rewritten as:
\beq
\langle E^2\rangle = \langle (T-\mu)^2\rangle + \langle V^2\rangle,
\eeq
where the first term can be estimated as follows:
\beq
\langle (T-\mu)^2\rangle = \mu k_{\eta}^2\left ( \alpha + 2\beta \alpha \frac{k_{\eta}}{k_{F}} + 
                                          \beta^{2}\left ( \frac{k_{\eta}}{k_{F}} \right )^{2} \right ) ,
\eeq
with $\alpha, \beta $ being numerical factors of the order of unity.
In order to estimate $\langle V^2\rangle$ 
let us assume now that $\Delta ({\bf r}) = \Delta_{out} + \delta\Delta ({\bf r})$  
and $\delta\Delta ({\bf r}) \ne 0$ for $r<R$. If we treat $\delta\Delta ({\bf r})$
as a small perturbation on top of the homogeneous pairing field $\Delta_{out}$ then:
\bea
\langle V^2\rangle &\approx& \Delta_{out}^{2} + 2\Delta_{out}
\int_{r<R} d^{3} r \chi^{T}({\bf r}) \delta\Delta ({\bf r}) \chi({\bf r}) = \nn \\
&=&\Delta_{out}^{2} + 2\Delta_{out} \delta\Delta
\int_{r<R} d^{3} r \chi^{T}({\bf r}) \chi({\bf r}),
\eea
where $\delta\Delta=\delta\Delta ({\bf r}_{0})$ and $r_{0}<R$.
Clearly, the last integral contains a smooth and a rapidly oscillating part.
The smooth part can be extracted using the mean value theorem:
\beq
\int_{r<R} d^{3} r \chi^{T}({\bf r}) \chi({\bf r}) = 
 \psi_{k_{\eta}}^{T} ({\bf r}_{1})\psi_{k_{\eta}}({\bf r}_{1}) 
 \int_{r<R} d^{3} r(\phi_{k_{F}}({\bf r}))^{2},
\eeq
where $r_{1}<R$. If $k_{F}R \gg 1$ then the right hand side integral is proportional to $R$
instead of $R^{3}$ as one would expect. This fact can be understood in the classical
picture since the probability of finding a free particle inside some area is proportional
to its linear size. On the other hand, since the above integral is dimensionless it should be
a function of $k_{\eta} R$. Hence one may write:
\beq
\langle V^2\rangle\approx\Delta_{out}^{2} + 2\Delta_{out} \delta\Delta f(k_{\eta} R),
\eeq
where $f(k_{\eta} R) \rightarrow \gamma k_{\eta}R$, if $k_{\eta}\rightarrow 0$ and
$\gamma >0$ is a numerical constant. Using the above estimates one may express
$\langle E^2\rangle$ as a function of $k_{\eta}$. Because we would like to answer the
question whether the minimum of energy corresponds to $k_{\eta}>0$, it is sufficient
if we consider the limit of $k_{\eta}$ being very small. In this case:
\beq
\langle E^2\rangle \approx  \mu k_{\eta}^2 \alpha^2
            + \Delta_{out}^{2} + 2\Delta_{out} \delta\Delta \gamma k_{\eta} R.
\eeq
Note the striking similarity with the estimation for the ground state energy
of a normal potential in one dimension.
The condition for the minimum of $\langle E^2\rangle$ gives the relation for
$k_{\eta}$:
\beq \label{cond}
k_{\eta}\approx -\frac{\gamma}{\alpha^{2}}\frac{\delta\Delta\Delta_{out} R}{\mu}.
\eeq
It shows that $\langle E^2\rangle$ is minimized by $k_{\eta}>0$, if 
$\Delta_{out} \delta\Delta < 0$. The last condition is obvious as it corresponds to
the pairing potential well, whereas a positive value would be equivalent to
a potential barrier. 

Hence the scattering on the pairing potential has a more severe effect on the 
quasiparticle spectrum as the pairing
field turns effectively the three-dimensional problem into a one-dimensional, 
and thus the scattering is more effective. 
\begin{figure}[tbh]
\epsfxsize=9.0cm
\centerline{\epsffile{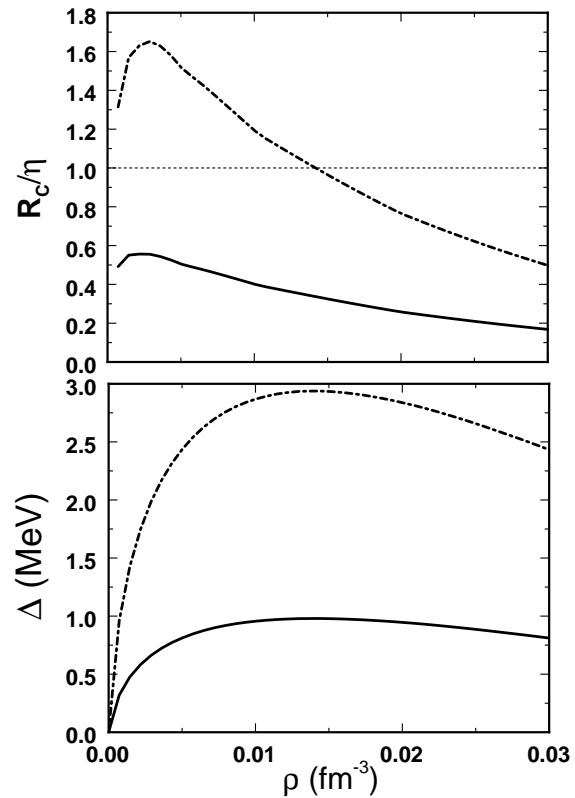}}
\caption{ 
The results of minimization of $\langle E^2\rangle$ with respect to 
localization length $\eta$ (see text) plotted as a function of the nuclear density in
the inner crust of neutron star. The dashed line corresponds to the BCS
pairing and the solid line denotes the pairing with inclusion of polarization
corrections. The lower subfigure shows
the coresponding pairing gaps, used in the calculations.}
\end{figure}

In order to quantitatively study this effect we minimize $ \langle E^2\rangle $
in a class of wave functions of the form:
\beq
\chi({\bf r}) = \frac{\sin(k_{F}r)}{k_{F}r}
                 \exp(-k_{\eta}^{2} r^{2} )
\left ( \begin{array}{l} u  \\   v
\end{array} \right )
\eeq
with $\displaystyle{\int d^{3} r \chi^{T}({\bf r}) \chi({\bf r}) =1 }$.
The pairing field was assumed to have the form:
\beq
\Delta (r) = \Delta_{out} + \delta\Delta \exp(-(r/\xi)^{2}),
\eeq
where $\delta \Delta=\Delta_{in}-\Delta_{out}$, and 
$\xi=\displaystyle{\frac{1}{k_{F}}\frac{\mu}{\Delta_{out}}}$.
The actual parameters $\Delta_{out}, \Delta_{in}$ requires some knowledge about
the density dependence of the pairing field. 
The two sets of calculation has been performed. In the first one, the
pairing field $\Delta_{out}$ was assumed to have the density dependence 
obtained from the BCS approach with Argonne v18 interaction\cite{kkc}, whereas
$\Delta_{in}$ corresponds to the neutron pairing at the nuclear saturation density.
The second set of calculation was performed for pairing extracted from
the s-wave N-N scattering data 
$\Delta \approx \displaystyle{\left (\frac{2}{e}\right )^{7/3}\frac{\hbar^{2}k_{F}^{2}}{2m}
\exp \left ( -\frac{\pi}{2\tan\delta (k_{F})}  \right)}$,
which is supposed to include the polarization corrections\cite{yu}.
The results are presented in the Fig. 1.
In the upper subfigure the $R_{C}/\eta $ is plotted, where $\eta=1/k_{\eta}$, and 
$R_{C}$ is the
Wigner-Seitz cell taken according to the calculations from refs. \cite{oya}.

The results indicate a strong dependence of $\eta$ on the pairing field. In the case of
stronger pairing the quasiparticle states become
practically localized for lower densities, i.e. $R_{C}/\eta > 1$. 
For weaker pairing the impact of the inhomogeneous
pairing field is less pronounced although the ratio $R_{C}/\eta $ reaches 0.6.
It has to be noted that the ratio 
$\displaystyle{ \sqrt{\langle E^{2}\rangle }/\Delta_{out} }\approx 0.9$ throughout
the crust. Consequently $\eta /\xi \approx 50$. 
It supports the conlcusion based on the schematic model.

Note that in the calculations the mean nuclear potential was set to zero.
Otherwise the results would have been obscured by the existence of an additional
scattering which would alter the rapidly oscillating part of the wave function. The current
approach corresponds to the Thomas-Fermi approximation for the mean-field
potential. Thus the results represent the average behavior of the
quasiparticle states around the Fermi level. The solution
of the full BdG or HFB equations
can change the above results in the following way:
around the Fermi level will appear resonances induced by the mean-field.
However their pairing properties will be practically unaltered, as compared
to our model, unless their energy will deviate from the Fermi level
more than $|\Delta_{out}|$.

The consequences of the presented results are twofold. First, they suggest that the
scattering of quasiparticles has to be taken into account if the neutron
band structure is to be determined. Consequently, for this purpose
the BCS approach is too poor, as there the quasiparticle scattering is neglected.
On the other hand BCS is sufficient if one is interested in the bulk properties
of the inner crust: energy density, density distribution, proton to neutron fraction, 
etc. It is interesting to note that the need for HFB (or BdG) approach has been
noticed in the context of nuclei close to the neutron drip line, long time
ago\cite{bul}. There the argument originated from the nuclear density
localization requirement.
Here, the main argument comes from the realization that the pairing field
exhibits spatial variations in the inhomogeneous nuclear matter. When
the matter becomes diluted the scattering on the pairing potential becomes
more effective and contribute to the band structure around the Fermi level.

Second, the consequence of neutron localization by the pairing field
may turn out to be important for transport properties. Namely, in this
case the effective mass has to be determined
and thus the band structure has to be calculated\cite{cham1}.
Within the approach we presented here, one may estimate the dispersion relation
for the band corresponding to bound neutrons.
Taking into account only the nearest neighbour interaction of 
the localized neutrons one obtains: 
$ E(k) \propto const. + \exp( -\frac{1}{2} (R_{C}/\eta)^{2} ) k^{2}$.
Note that in the limit of $R_{C}/\eta > 1$ the coefficient in front
is practically zero and the effective mass tends to infinity.
This corresponds to complete localization of neutron states at the
Fermi level around nuclear impurities. Clearly the detailed calculations
need to be performed in order to investigate this problem.

Summarizing, it was shown that the inhomogeneity of the pairing field
in the neutron matter leads to the formation of neutron bound states
around the Fermi level. 
They appear at an arbitrarily weak deviation from the homogeneous pairing field
and are the consequence of the focusing property of Andreev reflection. 

Discussions with Aurel Bulgac are 
gratefully acknowledged. This work was supported in part by the
Department of Energy under grant DE-FG03-97ER41014 and 
by the Polish Committee for Scientific Research
(KBN) under Contract No.~1~P03B~059~27. 

\end{document}